\documentclass{cambridge6A}
\usepackage{graphics}
\usepackage[active]{srcltx}
\usepackage{amsfonts}
\usepackage{amsmath}
\usepackage{bm}
\input{epsf}
\usepackage{epsfig,here}

\begin{document}

\chapter{Weak Values and Quantum Nonlocality}

\begin{center} \large Yakir Aharonov$^{1,2}$, Eliahu Cohen$^1$ \end{center}

 \begin{center} $^1$School of Physics and Astronomy, Tel Aviv University, Tel Aviv 69978, Israel, eliahuco@post.tau.ac.il \\
                $^2$Schmid College of Science, Chapman University, Orange, CA 92866, USA, yakir@post.tau.ac.il \end{center}

Entanglement and nonlocality are studied in the framework of pre-/post-selected ensembles with the aid of weak measurements and the Two-State-Vector Formalism.  In addition to the EPR-Bohm experiment, we revisit the Hardy and Cheshire Cat experiments, whose {\it entangled} pre- or post-selected states give rise to curious phenomena. We then turn to even more peculiar phenomenon suggesting ``emerging correlations'' between {\it independent} pre- and post-selected ensembles of particles. This can be viewed as a quantum violation of the classical ``pigeonhole principle''.

\section{Introduction}

It is seldom acknowledged that 7 years before the celebrated Bell paper \cite{Bell}, Bohm and Aharonov \cite{BH} published an analysis of the EPR paradox \cite{EPR}. They suggested an experimental setup, based on Compton scattering, for testing nonlocal correlations between the polarizations of two annihilation photons. In 1964, Bell proposed his general inequality thereby excluding local realism. During the same time, Aharonov {\it et al.} constructed the foundations of a time-symmetric formalism of quantum mechanics \cite{ABL}. While Bell's proof utilizes entanglement for demonstrating nonlocal correlations, we will describe in what follows the emergence of nonlocal correlations between product states. For this purpose, however, we shall invoke weak measurements of pre- and post-selected ensembles.

In classical mechanics, initial conditions of position and velocity for every particle fully determine the time evolution of the system. Therefore, trying to impose an additional final condition would either lead to redundancy or inconsistency with the initial conditions. This is radically different in quantum mechanics. Because of the uncertainty principle, an initial state-vector does not fully determine, in general, the outcome of a future measurement. However, adding a final (backward-evolving) state-vector, results in a more complete description of the quantum system in-between these two boundary conditions, that has bearings on the determination of measurement outcomes.

The basis for this time-symmetric formulation of quantum mechanics was laid by Aharonov, Bergman, and Lebowitz (ABL), who derived a symmetric probability rule concerning measurements performed on systems, while taking into account the final state of the system, in addition to the usual initial state \cite{ABL}. Such a final state may arise due to a post-selection, that is, performing an additional measurement on the system and considering only the cases with the desired outcome. Since then, the time-symmetric formalism was further generalized (see for instance \cite{TSVF,Reznik}) and was shown to be very helpful for understanding conceptual ideas in quantum mechanics, such as the past of the quantum particle \cite{Past}, the measurement problem \cite{Collapse} and more.

In order to verify the two-state description without intervening with the final (post-selected) boundary condition, a subtle kind of quantum measurement was suggested- {\it weak measurement} \cite{AAV}. Weak measurements are based on the von Neumann scheme for performing quantum measurements, albeit with a very small coupling compared to the measurement's uncertainty. The weak coupling created between the measured system and the measuring (quantum) pointer does not change significantly the measured state, yet provides robust information when an ensemble of states in discussed \cite{AAV,ACE}. Given an operator $A$ we wish to measure on a system $|\psi\rangle$, the coupling to the measuring pointer is achieved through the Hamiltonian
\begin{equation}\label{Hint}
H_{int}=\epsilon g(t)AP_d
\end{equation}
where $\epsilon<<1$ is a small parameter, $\int_0^T g(t)dt = 1$ for a measurement of duration $T$ and $P_d$ is the pointer's momentum. The result of this coupling to a pre- and post-selected ensemble $\langle \phi|~|\psi\rangle$ ({\it i.e.} the reading of the pointer) is known as a {\it weak value} \cite{AAV}:
\begin{equation}\label{WV}
\langle A \rangle_w=\frac{\langle \phi|A|\psi\rangle}{\langle \phi|\psi\rangle}
\end{equation}
Weak measurements were shown to be very useful in analyzing a variety of problems \cite{SpinHall,SNR,WhiteLight,Jordan}. We will focus henceforth on entanglement and nonlocality.

The outline of the paper is as follows: Section 1 describes 3 experiments with entangled pre- or post-selected states: Hardy's paradox, the Cheshire Cat, and finally an EPR-Bohm experiment. Section 2 presents the analysis of ``emerging correlations'' within non-entangled pre- and post-selected system.

\section{Entangled Pre- and Post-Selected Systems} \label{sec2}
We shall revisit 3 gedanken experiments which highlight the unique features of weak values between entangled pre- and post-selected states.
\subsection{Hardy's Experiment}
\begin{figure}[h]
 \centering \includegraphics[height=6cm]{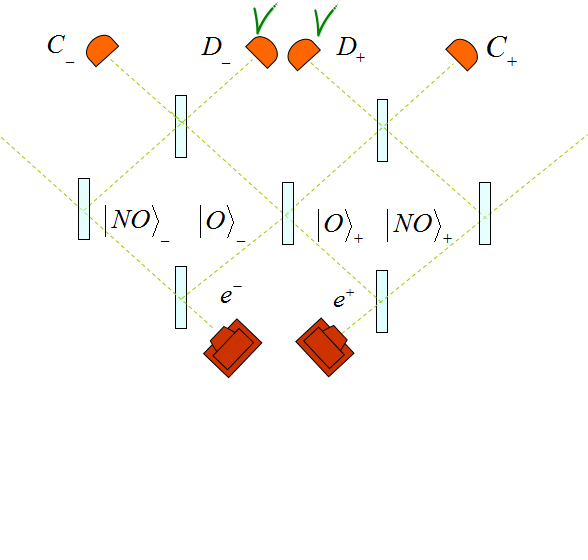}
      \caption{{\bf Hardy's experiment}} \label{Hardy}
\end{figure}

An interesting demonstration of weak values between an entangled pre-selected state and a product post-selected state, as well as a conceptual success of the TSVF, is given by the Hardy experiment \cite{Hardy,WHardy}. Two Mach-Zehnder interferometers overlap in one corner (See Fig. \ref{Hardy}). Their length is tuned such that electron entering the first will always arrive at detector $C_-$ while a positron entering the second will always arrive at detector $C_+$.  Hence, when an electron and a positron simultaneously traverse the setup, they might annihilate or make their partner reach the ``forbidden'' detector $D_- / D_+$.  In case no annihilation was recorded we know that the state of the particles is
\begin{equation} |\psi_i\rangle=\frac{1}{\sqrt{3}}[|O\rangle_+|NO\rangle_-+|NO\rangle_+|O\rangle_-+|NO\rangle_+|NO\rangle_-],
\end{equation}
{\it i.e.}, at least one of the particles took the non-overlapping ($NO$) state, thereby excluding the case they both took the overlapping path ($O$).
The interferometers were tuned such that $C_-$ clicks for the $\frac{1}{\sqrt{2}}(|O\rangle_-+|NO\rangle_-)$ state, $D_-$ clicks for the $\frac{1}{\sqrt{2}}(|O\rangle_- - |NO\rangle_-)$ state, and similarly for $C_+$ and $D_+$.
Therefore, choosing the case of clicks at $D_-$ and $D_+$ amounts to post-selection of the state
\begin{equation}
|\psi_f\rangle=\frac{1}{2}(|O\rangle_+ - |NO\rangle_+) (|O\rangle_- - |NO\rangle_-).
\end{equation}
This post-selection is possible, because $\psi_i$ is not orthogonal to $\psi_f$, but it is peculiar nevertheless: detection of the electron at $D_-$ naively tells us that the positron took its overlapping path, while detection of the position at $D_+$ naively tells us that the electron took its overlapping path. This scenario, however, is impossible, because we know annihilation did not take place. The paradox is resolved using the TSVF. When we calculate the weak values of the various projection operators we find out that
\begin{equation}
\langle\Pi_{O}^- \Pi_{O}^+\rangle_w=0
\end{equation}
and
\begin{equation}
\langle\Pi_{NO}^- \Pi_{O}^+\rangle_w=\langle\Pi_{O}^- \Pi_{NO}^+\rangle_w=+1,
\end{equation}
while
\begin{equation}
\langle\Pi_{NO}^- \Pi_{NO}^+\rangle_w=-1.
\end{equation}
This leads us to conclude that although the number of pairs is 1, we have two ``positive'' pairs and one ``negative'' pair- a pair of particles with opposite properties. The pair in the ``NO-NO'' path creates a negative ``weak potential'' \cite{Potential}, that is, when weakly interacting with any other particle in the intermediate time, its effect will have a negative sign.

\subsection{The Cheshire Cat}
The second demonstration is the ``Cheshire Cat'' \cite{Cheshire}. Let a particle (the ``Cat'') have two degrees of freedom: spatial $|L\rangle$,$|R\rangle$ (the cat is on the left/right box) and spinorial $|\uparrow\rangle$,$|\downarrow\rangle$ (the Cat is smiling/frowning ). The Cat is pre-selected at $t=0$ in the entangled state
\begin{equation}
|\psi_i\rangle=\frac{1}{2}(|\uparrow\rangle+|\downarrow\rangle)|L\rangle+ \frac{1}{2}(|\uparrow\rangle-|\downarrow\rangle) |R\rangle,
\end{equation}

and post-selected at $t=T$ in the product state
\begin{equation}
|\psi_f\rangle=\frac{1}{2}(|\uparrow\rangle-|\downarrow\rangle) (|L\rangle+|R\rangle).
\end{equation}
At $0<t<T$ the Cat is in the right box, since:
\begin{equation}
\langle \Pi_L \rangle_w=0~,~\langle \Pi_R \rangle_w=1
\end{equation}
and it is smiling, since
\begin{equation}
\langle \sigma_z \rangle_w =1,
\end{equation}
but its smile is in the left box (!) since
\begin{equation}
\langle \sigma_z \Pi_L \rangle_w=1~,~\langle \sigma_z \Pi_R \rangle_w=0.
\end{equation}
This can be understood as the failure of the product rule for weak values between pre- and post-selected states. Weak values reveal a perplexing phenomenon: the spin of a quantum particle can be separated from it mass.

\subsection{An EPR-Bohm Experiment}
The third demonstration is an EPR-Bohm experiment \cite{AR} (for a GHZ-like demonstration, where a set of $N$ particles is faced with the four GHZ mutual-exclusive requirements see \cite{Peculiar}). Alice and Bob share an ensemble of spin-$1/2$ entangled particles prepared in:
\begin{equation}
|\psi_i\rangle=\frac{1}{\sqrt{2}}(|\uparrow\rangle_A|\downarrow\rangle_B-|\downarrow\rangle_A|\uparrow\rangle_B).
\end{equation}
Alice and Bob measure their particles along axes that they choose at random from a finite set. Suppose Alice measures her spin along the $x$-axis and Bob measures his spin along the $y$-axis, and the outcomes are
\begin{equation}
|\sigma_x\rangle_A=1,~,|\sigma_y\rangle_B=1,
\end{equation}
{\it i.e.}, the post-selected state is
\begin{equation}
|\psi_f\rangle=\frac{1}{2}(|\uparrow\rangle_A+|\downarrow\rangle_A)(|\uparrow\rangle_B+i|\downarrow\rangle_B).
\end{equation}
According to the EPR paradox, the results of Alice's measurement cannot depend on Bob's choice of axes and vice-versa. Therefore, $\sigma_x^A$, $\sigma_y^A$, $\sigma_x^B$, $\sigma_y^B$ are all elements of reality (in the EPR sense).
We now note that $|\psi_i\rangle$ is an eigenvalue of the three operators $\sigma_x^A\sigma_x^B$, $\sigma_y^A\sigma_y^B$ (with eigenvalue $-1$) and $\sigma_x^A\sigma_y^B+\sigma_y^A\sigma_x^B$ (with eigenvalue $0$). Therefore, the post-selection accords on the one hand with $\sigma_x^B=\sigma_y^A=-1$, but on the other hand, with $\sigma_y^A\sigma_x^B=-1$.~An apparent contradiction!
To resolve this paradox we turn again to the weak values of the various projection operators.
\begin{equation}
\begin{array} {lcl}
\langle\Pi_{\uparrow y}^A\Pi_{\uparrow x}^B\rangle_w=-1/2 \\
\langle\Pi_{\uparrow y}^A\Pi_{\downarrow x}^B\rangle_w=1/2 \\
\langle\Pi_{\downarrow y}^A\Pi_{\uparrow x}^B\rangle_w=1/2 \\
\langle\Pi_{\downarrow y}^A\Pi_{\downarrow x}^B\rangle_w=1/2.
\end{array}
\end{equation}
Hence, for Alice's system:
\begin{equation}
\begin{array} {lcl}
\langle\Pi_{\uparrow y}^A\rangle_w=\langle\Pi_{\uparrow y}^A\Pi_{\uparrow x}^B\rangle_w +\langle\Pi_{\uparrow y}^A\Pi_{\downarrow x}^B\rangle_w =0\\
\langle\Pi_{\downarrow y}^A\rangle_w=\langle\Pi_{\downarrow y}^A\Pi_{\uparrow x}^B\rangle_w+\langle\Pi_{\downarrow y}^A\Pi_{\downarrow x}^B\rangle_w=1,
\end{array}
\end{equation}
consistent with the requirement $\sigma_y^A=-1$. Similarly for Bob,
\begin{equation}
\begin{array} {lcl}
\langle\Pi_{\uparrow x}^B\rangle_w=\langle\Pi_{\uparrow y}^A\Pi_{\uparrow x}^B\rangle_w +\langle\Pi_{\downarrow y}^A\Pi_{\uparrow x}^B\rangle_w =0 \\
\langle\Pi_{\downarrow x}^B\rangle_w=\langle\Pi_{\uparrow y}^A\Pi_{\downarrow x}^B\rangle_w+\langle\Pi_{\downarrow y}^A\Pi_{\downarrow x}^B\rangle_w=1,
\end{array}
\end{equation}
consistent with $\sigma_x^B=-1$.
In addition
\begin{equation}
\begin{array} {lcl}
\langle\sigma_y^A\sigma_x^B\rangle_w=\langle\Pi_{\uparrow y}^A\Pi_{\uparrow x}^B\rangle_w-\langle\Pi_{\uparrow y}^A\Pi_{\downarrow x}^B\rangle_w-\langle\Pi_{\downarrow y}^A\Pi_{\uparrow x}^B\rangle_w+\langle\Pi_{\downarrow y}^A\Pi_{\downarrow x}^B\rangle_w=-1,
\end{array}
\end{equation}
consistent with $\sigma_y^A\sigma_x^B=-1$.

Hence we see an alternative way of understanding quantum nonlocality. The classical limitation on correlations can be violated by quantum weak values which are negative. Like the case of Hardy's experiment, these weak values should be understood as reversing the interaction sign, rather than as negative probabilities. In fact, each weak value (not necessarily a peculiar one) defines a ``weak potential'' within a pre-/post-selected ensemble \cite{Potential}.

For a more complex setup of an EPR-Bohm experiment with weak measurements see \cite{Future}.

\section{Non-Entangled Pre- and Post-Selected Ensembles}

\begin{figure}[h]
 \centering \includegraphics[height=6cm]{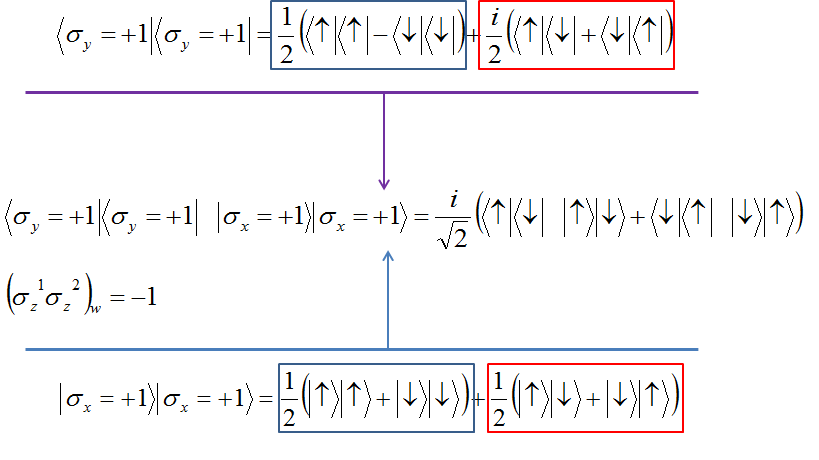}
      \caption{{\bf Emerging correlations between independent ensembles}} \label{Corr}
\end{figure}

Let two spins be independently prepared at $t=0$ in the state $|\sigma_x=+1\rangle$, to create the product state
\begin{equation}
|\sigma_x=+1\rangle_1|\sigma_x=+1\rangle_2.
\end{equation}
Suppose that later, at time $t=T$ they are independently measured along the $y$-axis and found at:
\begin{equation}
|\sigma_y=+1\rangle_1|\sigma_y=+1\rangle_2.
\end{equation}
Could there be correlations between these two independent spins at times $0<t<T$?\\
The correlation between operators $A$ and $B$ is defined according to
\begin{equation}
Corr(A,B) \equiv \langle AB \rangle - \langle A \rangle \langle B \rangle.
\end{equation}
If measured strongly, $\sigma_z$ between $t=0$ and $t=T$ would be clearly found to have a zero expectation value, $\langle \sigma_z \rangle=0$, for both particles, since they were prepared in an eigenstate of $\sigma_x$ and post-selected in an eigenstate of $\sigma_y$.
But what is the product of their spins along the $z$-axis? Eq. \ref{WV} tells us that the weak value of $\sigma_z$ is
\begin{equation}
\langle \sigma_z \rangle_w=i.
\end{equation}
We saw in Sec. \ref{sec2} the breakdown of the product rule for entangled states, but now the pre- and post-selected states are not entangled and hence
\begin{equation}
\langle \sigma_z^{(1)} \sigma_z^{(2)} \rangle_w=i\cdot i= -1.
\end{equation}
In addition, for dichotomic operators we know that if the weak value equals one of the eigenvalues, then it also equals the strong value. Hence,
\begin{equation}
\langle \sigma_z^{(1)} \sigma_z^{(2)} \rangle= -1,
\end{equation}
and
\begin{equation}
Corr(\sigma_z^{(1)}, \sigma_z^{(2)})= -1.
\end{equation}
We thus see that the two spins were anti-correlated along the $z$-axis, but they were also correlated along the $x$-axis and along the $y$-axis, so they must have been maximally entangled. But alas, they were pre- and post-selected in a product state!
To better understand why these particles seem to be maximally entangled, we can represent their initial and final states in the $z$ basis:
\begin{equation}
|\sigma_x=+1\rangle_1|\sigma_x=+1\rangle_2=\frac{1}{2}(|\uparrow \uparrow\rangle+|\uparrow \downarrow\rangle+|\downarrow \uparrow\rangle+|\downarrow \downarrow\rangle),
\end{equation}
and
\begin{equation}
|\sigma_y=+1\rangle_1|\sigma_y=+1\rangle_2=\frac{1}{2}(|\uparrow \uparrow\rangle+i|\uparrow \downarrow\rangle+i|\downarrow \uparrow\rangle-|\downarrow \downarrow\rangle).
\end{equation}
Hence, the correlated part in the pre- and post-selected states cancels due to orthogonality, and only the anti-correlated part remains (See Fig. \ref{Corr}). These correlations can be verified, for example, by performing nonlocal measurements \cite{NonL1,NonL2}.
This, in fact, is a very general phenomenon occurring each time the pre- and post-selected states do not coincide. In these cases the states will have orthogonal parts whose cancelation would yield correlations.
Repeating this procedure for an ensemble of $N$ particles, we find each pair to be maximally entangled in an apparent violation of ``entanglement monogamy''.  However, this entanglement is of subtle kind since both the pre- and post-selected states were not entangled in the first place. Moreover, it cannot be verified on each particle alone (only on pairs) and cannot be used for teleportation.
It turns out the weak values contain, in some sense, even more information. In the above scenario the two experimenters need not know what are the pre- and post-selected outcome, they just need to know which particles had the same outcomes, and then performing weak measurements along the $x$,$y$ and $z$ axes they would know which direction was chosen and which outcome was measured. Indeed, it can be shown that
\begin{equation}
(\sigma_x)_w^2+(\sigma_y)_w^2+(\sigma_z)_w^2=1,
\end{equation}
for any pre- and post-selected ensemble.
We thus understand that quantum correlations underlie almost any experiment, but are only visible upon post-selection and grouping of similar results. \\
This gedanken experiment can be viewed as demonstrating the breakdown of the classical ``Pigeonhole principle''. It was previously shown \cite{Pigeonhole} that special pre- and post-selection of a quantum system lead to unusual correlations between its parts. Here we witness once more the appearance of emerging correlations in a pre-/post-selected quantum ensemble. Thinking about $\sigma_z$ as denoting the position of a particle in one of two boxes, we can see that within a group of 3 particles with the above pre- and post-selection, every pair is anti-correlated, that is, no pair resides in the same box. This clearly stands in contrast with the classical principle, according to which at least one pair of pigeons within a group of 3 pigeons must share the same hole. The result can be trivially generalized to $N$ particles.

\section{Discussion}
Bell's proof inclined us to think of entangled states as concealing nonlocal correlations. Within pre- and post-selected ensembles, these correlations are responsible for intriguing effects. Yet the truly curious result we have just seen is the emergence of nonlocal correlations in practically every pre- and post-selected ensemble of product states.
\\
Weak measurements were demonstrated once more to provide us with a richer description of the quantum reality. Negative weak values were shown to be essential for understanding both the Hardy, Cheshire Cat and EPR-Bohm experiments, while imaginary weak values indicated emerging correlations in a product state. These results accord well with a previous work of Marcovitch, Reznik and Vaidman \cite{MRV}, where correlations within pre- and post-selected ensembles were shown to exceed Tsirelson's bound \cite{Tsi} and reach Popescu-Rohrlich bound \cite{PR}.
\\
We feel that the current research is not over yet. Weak values between entangled states might have even a more crucial role in understanding fundamental questions such as the information paradox in black holes \cite{HM} and time \cite{Each}.

\hfill

{\bf Acknowledgements}
\\ This work has been supported in part by the Israel Science Foundation Grant No. 1311/14. We would like to thank Aharon Brodutch and Tomer Landsberger for helpful comments and discussions.


\begin{thebibliography} {22}

\bibitem{Bell}
J.S. Bell, On the einstein-podolsky-rosen paradox, Physics {\bf  1}, 195-200 (1964).

\bibitem{BH}
D. Bohm, Y. Aharonov, Discussion of Experimental Proof for the Paradox of Einstein, Rosen, and Podolsky, Phys. Rev. {\bf 108}, 1070-1076 (1957).

\bibitem{EPR}
A. Einstein, B. Podolsky, N. Rosen, Can quantum-mechanical description of physical reality be considered complete, Phys. Rev. {\bf 47}, 777-280 (1935).

\bibitem{ABL}
Y. Aharonov, P.G. Bergmann, J.L. Lebowitz, Time symmetry in the quantum process of measurement, Phys. Rev. B. {\bf 134}, 1410-1416 (1964).


\bibitem{TSVF}
Y. Aharonov and L. Vaidman,
The Two-State Vector Formalism of Quantum Mechanics, in {\it Time in Quantum Mechanics}, J.G. Muga {\it et al.} eds., Springer, 369-412 (2002).


\bibitem{Reznik}
B. Reznik and Y. Aharonov, Time-symmetric formulation of quantum mechanics, Phys. Rev. A {\bf 52}, 2538-2550 (1995).


\bibitem{Past}
L. Vaidman, Past of a quantum particle, Phys. Rev. A {\bf 87}, 052104 (2013).

\bibitem{Collapse}
Y. Aharonov, E. Cohen, E. Gruss, T. Landsberger, Measurement and collapse within the two-state-vector formalism, Quantum Stud.: Math. Found. {\bf 1} , 133-146 (2014).

\bibitem{AAV}
Y. Aharonov,  D.Z. Albert, L. Vaidman,  How the result of a measurement of a component of the spin of a spin-1/2 particle can turn out to be 100, Phys. Rev. Lett. {\bf  60}, 1351-1354 (1988).

\bibitem{ACE}
Y. Aharonov, E. Cohen, A.C. Elitzur, Foundations and applications of weak quantum measurements, Phys. Rev. A {\bf 89}, 052105 (2014).

\bibitem{SpinHall}
O. Hosten, P. Kwiat Observation of the spin Hall effect of light via weak measurements, Science {\bf 319}, 787-790 (2008).

\bibitem{SNR}
D.J. Starling,  P.B. Dixon, A.N. Jordan, J.C. Howell, Optimizing the signal-to-noise ratio of a beam-deflection measurement with interferometric weak values, Phys. Rev. A {\bf 80}, 041803 (2009).

\bibitem{WhiteLight}
X.Y Xu, Y. Kedem, K. Sun, L. Vaidman, C.F. Li, G.C. Guo, Phase estimation with weak measurement using a white light source, Phys. Rev. Lett. {\bf 111}, 033604 (2013).

\bibitem{Jordan}
A.N Jordan, J. Tollaksen, J.E. Troupe, J. Dressel, Y. Aharonov, Heisenberg scaling with weak measurement: A quantum state discrimination point of view, arXiv:1409.3488 (2014).

\bibitem{Hardy}
L. Hardy, Quantum mechanics, local realistic theories, and Lorentz-invariant realistic theories, Phys. Rev. Lett., {\bf  68}, 2981 (1992).

\bibitem{WHardy}
Y. Aharonov, A. Botero, S. Popescu, B. Reznik, J. Tollaksen, Revisiting Hardy's paradox: counterfactual statements, real measurements, entanglement and weak values, Phys. Lett. A {\bf 301}, 130-138 (2002).

\bibitem{Potential}
Y. Aharonov,  E. Cohen, S. Ben-Moshe, Unusual interactions of pre- and past-selected particles, EPJ Web Conf. {\bf 70}, 00053 (2014).

\bibitem{Cheshire}
Y. Aharonov, S. Popescu, D. Rohrlich, P. Skrzypczyk, Quantum Cheshire cats, New J. Phys. {\bf 15}, 113015 (2013).

\bibitem{AR}
Y. Aharonov, D. Rohrlich, Quantum Paradoxes: Quantum theory for the perplexed, Weinheim: Wiley-VCH, Ch. 17, (2005).

\bibitem{Peculiar}
Y. Aharonov, S. Nussinov, S. Popescu, and L. Vaidman, Peculiar features of entangled states with postselection,
Phys. Rev. A {\bf 87}, 014105 (2012).

\bibitem{Future}
Y. Aharonov, E. Cohen, D. Grossman, A.C. Elitzur, Can Weak Measurement Lend Empirical Support to Quantum Retrocausality?, EPJ Web Conf. {\bf 58} (2013).

\bibitem{NonL1}
Y. Aharonov, D. Z. Albert, L. Vaidman, Measurement process in relativistic quantum theory, Phys. Rev. D {\bf 34}, 1805-1813 (1986).

\bibitem{NonL2}
A. Brodutch, E. Cohen, Weak measurements via quantum erasure, arXiv:1409.1575 (2014).

\bibitem{Pigeonhole}
Y. Aharonov, F. Colombo, S. Popescu, I. Sabadini, D.C.Struppa, J. Tollaksen, The quantum pigeonhole principle and the nature of quantum correlations, arXiv:1407.3194 (2014).

\bibitem{MRV}
S. Marcovitch, B. Reznik, L. Vaidman, Quantum-mechanical realization of a Popescu-Rohrlich box, Phys. Rev. A {\bf 75}, 022102 (2007).

\bibitem{Tsi}
B.S. Cirel'son, Quantum generalizations of Bell's inequality, Lett. Math. Phys. {\bf 4}, 93-100 (1980).

\bibitem{PR}
S. Popescu, D. Rohrlich, Quantum nonlocality as an axiom, Found. Phys. {\bf 24}, 379-385 (1994).

\bibitem{HM}
G. Horowitz, J. Maldacena,  The black hole final state, J. High Energy Phys. {\bf 2004.02}, 008 (2004).

\bibitem{Each}
Y. Aharonov, S. Popescu, J. Tollaksen,  Each instant of time a new Universe, Quantum Theory: A Two-Time Success Story. Springer-Milan,  21-36 (2014).

\end{thebibliography}
\end{document}